\documentclass[submission,copyright,creativecommons]{eptcs}
\providecommand{\event}{PROLE 2014} 

\usepackage{makeidx}
\usepackage{graphicx,latexsym,alltt}
\usepackage{amssymb,amsmath}
\usepackage{algorithm,algorithmic}
\usepackage{subfigure}
\usepackage{color}
\usepackage{url}
\usepackage[latin1]{inputenc}

\newcommand{\figures}[1]{./Figures/#1}
\newcommand{\biblio}[1]{./Biblio/#1}

\definecolor{blank}{rgb}{0.7,0.7,0.7}

\newtheorem{theorem}{Theorem}[section]
\newtheorem{definition}[theorem]{Definition}
\newtheorem{example}[theorem]{Example}

\long\def\comment#1{}

\renewcommand{\phi}{\varphi}

\def\defemb#1#2{\expandafter\def\csname #1\endcsname
                              {\relax\ifmmode #2\else\hbox{$#2$}\fi}}
\defemb{cA}{{\cal A}}
\defemb{cB}{{\cal B}}
\defemb{cC}{{\cal C}}
\defemb{cD}{{\cal D}}
\defemb{cE}{{\cal E}}
\defemb{cF}{{\cal F}}
\defemb{cG}{{\cal G}}
\defemb{cH}{{\cal H}}
\defemb{cI}{{\cal I}}
\defemb{cJ}{{\cal J}}
\defemb{cL}{{\cal L}}
\defemb{cM}{{\cal M}}
\defemb{cN}{{\cal N}}
\defemb{cO}{{\cal O}}
\defemb{cP}{{\cal P}}
\defemb{cR}{{\cal R}}
\defemb{cS}{{\cal S}}
\defemb{cT}{{\cal T}}
\defemb{cU}{{\cal U}}
\defemb{cV}{{\cal V}}
\defemb{cW}{{\cal W}}
\defemb{cX}{{\cal X}}
\defemb{cZ}{{\cal Z}}

\makeatletter
\newenvironment{prog}{\vspace{1.0ex}\par
\obeylines\@vobeyspaces\tt}{\vspace{1.0ex}\noindent
}
\makeatother
\newcommand{\startprog}{\begin{prog}}
\newcommand{\stopprog}{\end{prog}\noindent}

\catcode`\_=\active
\let_=\sb
\catcode`\_=12
\makeatother

\title{Web Template Extraction Based on Hyperlink Analysis} 

\author{
Juli\'an Alarte  \qquad\qquad
David Insa  \qquad\qquad 
Josep Silva
\institute{
Departamento de Sistemas Inform\'aticos y Computaci\'on\\
Universitat Polit\`ecnica de Val\`encia, Valencia, Spain
}
\email{ jualal@doctor.upv.es \qquad dinsa@dsic.upv.es \qquad jsilva@dsic.upv.es}
\and
Salvador Tamarit
\institute{
Babel Research Group\\
Universidad Polit\'ecnica de Madrid, Madrid, Spain
}
\email{stamarit@babel.ls.fi.upm.es}
}
\def\titlerunning{Web Template Extraction Based on Hyperlink Analysis}
\def\authorrunning{J. Alarte, D. Insa, J. Silva \& S. Tamarit}

\begin{document}

\maketitle


\begin{abstract}
Web templates are one of the main development resources for website engineers. 
Templates allow them to increase productivity by plugin content into already formatted and prepared pagelets. For the final user templates are also useful, because they provide uniformity and a common look and feel for all webpages.   
However, from the point of view of crawlers and indexers, templates are an important problem, because templates usually contain irrelevant information such as advertisements, menus, and banners. Processing and storing this information is likely to lead to a waste of resources (storage space, bandwidth, etc.).
It has been measured that templates represent between 40\% and 50\% of data on the Web. Therefore, identifying templates is essential for indexing tasks. In this work we propose a novel method for automatic template extraction that is based on similarity analysis between the DOM trees of a collection of webpages that are detected using menus information. Our implementation and experiments demonstrate the usefulness of the technique.
\end{abstract}

\section{Introduction}

A web template (in the following just template) is a prepared HTML page where formatting is already implemented and visual components are ready so that we can insert content into them. 
Templates are used as a basis for composing new webpages that share a common look and feel. This is good for web development because many tasks can be automated thanks to the reuse of components. In fact, many websites are maintained automatically by code generators that generate webpages using templates. Templates are also good for users, which can benefit from intuitive and uniform designs with a common vocabulary of colored and formatted visual elements.  

Templates are also important for crawlers and indexers, because they usually judge the relevance of a webpage according to the frequency and distribution of terms and hyperlinks. 
Since templates contain a considerable number of
common terms and hyperlinks that are replicated in a large
number of webpages, relevance may turn out to be inaccurate, leading to incorrect results (see, e.g., \cite{BarR02,VieSPMCF06,YiLL03}). 
Moreover, in general, templates do not contain relevant content, they usually contain one or more pagelets \cite{Cha01,BarR02} (i.e., self-contained logical regions with a well defined topic or functionality) where the main content must be inserted. 
Therefore, detecting templates can allow indexers to identify the main content of the webpage. 

Modern crawlers and indexers do not treat all terms in a webpage in the same way. Webpages are preprocessed to identify the template because template extraction allows them to identify those pagelets that only contain noisy information such as advertisements and banners. This content should not be indexed in the same way as the relevant content. Indexing the non-content part of templates not only affects accuracy, it also affects performance and can lead to a waste of storage space, bandwidth, and time.

Template extraction helps indexers to isolate the main content. 
This allows us to enhance indexers by assigning higher weights to the really relevant terms.
Once templates have been extracted, they are processed for indexing---they can be analyzed only once for all webpages using the same template---. Moreover, links in templates allow indexers to discover the topology of a website (e.g., through navigational content such as menus), thus identifying the main webpages. They are also essential to compute pageranks.

Gibson et al. \cite{GibPT05} determined that templates represent between 40\% and 50\% of data on the Web and that around 30\% of the visible terms and hyperlinks appear in templates. This justifies the importance of template removal \cite{YiLL03,VieSPMCF06} for web mining and search. 

Our approach to template extraction is based on the DOM \cite{DOM} structures that represent webpages. 
Roughly, given a webpage in a website, we first identify a set of webpages that are likely to share a template with it, and then, we analyze these webpages to identify the part of their DOM trees that is common with the original webpage. This slice of the DOM tree is returned as the template.  

Our technique introduces a new idea to automatically find a set of webpages that potentially share a template. Roughly, we detect the template's menu and analyze its links to identify a set of mutually linked webpages. 
One of the main functions of a template is in aiding navigation, thus almost all templates provide a large number of links, shared by all webpages implementing the template. Locating the menu allows us to identify in the topology of the website the main webpages of each category or section. These webpages very likely share the same template. 
This idea is simple but powerful and, contrarily to other approaches, it allows the technique to only analyze a reduced set of webpages to identify the template.

The rest of the paper has been structured as follows:
In Section~\ref{sec_rel} we discuss the state of the art and show some problems of current techniques that can be solved with our approach. 
In Section~\ref{sec_prelim} we provide some preliminary definitions and useful notation. 
Then, in Section~\ref{sec_temex}, we present our technique with examples and explain the algorithms used. 
In Section~\ref{sec_impl} we give some details about the implementation and show the results obtained from a collection of benchmarks.
Finally, Section~\ref{sec_concl} concludes.

\section{Related Work}\label{sec_rel}

Template detection and extraction are hot topics due to their direct application to web mining, searching, indexing, and web development. For this reason, there are many approaches that try to face this problem.
Some of them are especially thought for boilerplate removal and content extraction; and they have been presented in the CleanEval competition \cite{BarCKS08}, which proposes a collection of examples to be analyzed with a gold standard.

\emph{Content Extraction} is a discipline very close to template extraction. Content extraction tries to isolate the pagelet with the main content of the webpage. It is an instance of a more general discipline called \emph{Block Detection} that tries to isolate every pagelet in a webpage. There are many works in these fields (see, e.g., \cite{Got08,WenHH10,CarJLRC11,InsST13}), and all of them are directly related to template extraction. 

In the area of template extraction, there are three different ways to solve the problem, namely, (i) using the textual information of the webpage (i.e., the HTML code), (ii) using the rendered image of the webpage in the browser, and (iii) using the DOM tree of the webpage. 

The first approach analyzes the plain HTML code, and it is based on the idea that the main content of the webpage has more density of text, with less labels. For instance, the main content can be identified selecting the largest contiguous text area with
the least amount of HTML tags \cite{FerZBB08}. This has been measured directly on the HTML code by counting the number of characters inside text, and characters inside labels. This measure produces a ratio called CETR \cite{WenHH10} used to discriminate the main content. Other approaches exploit densitometric features based on the observation that some specific terms are more common in templates \cite{KohN08,Koh09}. 
The distribution of the code between the lines of a webpage is not necessarily the one expected by the user. The format of the HTML code can be completely unbalanced (i.e., without tabulations, spaces or even carriage returns), specially when it is generated by a non-human directed system. As a common example, the reader can see the source code of the main Google's webpage. At the time of writing these lines, all the code of the webpage is distributed in only a few lines without any legible structure. In this kind of webpages CETR is useless.

The second approach assumes that the main content of a webpage is often located in the central part and (at least partially) visible without scrolling \cite{BurR09}. This approach has been less studied because rendering webpages for classification is a computational expensive operation \cite{KohFN10}. 

The third approach is where our technique falls. While some works try to identify pagelets analyzing the DOM tree with heuristics \cite{BarR02}, others try to find common subtrees in the DOM trees of a collection of webpages in the website \cite{YiLL03,VieSPMCF06}. Our technique is similar to these last two works.  

Even though \cite{YiLL03} uses a method for template extraction, its main goal is to remove redundant parts of a website. 
For this, they use the Site Style Tree (SST), a data structure that is constructed by analyzing a set of DOM trees and recording every node found, so that repeated nodes are identified by using counters in the SST nodes. Hence, an SST summarizes a set of DOM trees. After the SST is built, they have information about the repetition of nodes. The most repeated nodes are more likely to belong to a noisy part that is removed from the webpages. Unfortunately, this approach does not use any method to detect the webpages that share the same template. They are randomly selected, and this can negatively influence the performance and the precision of the technique. 

In \cite{VieSPMCF06}, the approach is based on discovering optimal mappings between DOM trees. 
This mapping relates nodes that are considered redundant.
Their technique uses the RTDM-TD algorithm to compute a special kind of mapping called \emph{restricted top-down mapping} \cite{ReiGSL04}. 
Their objective, as ours, is template extraction, but there are two important differences. First, we compute another kind of mapping to identify redundant nodes. Our mapping is more restrictive because it forces all nodes that form pairs in the mapping to be equal. Second, in order to select the webpages of the website that should be mapped to identify the template, they pick random webpages until a threshold is reached. In their experiments, they approximated this threshold as a
few dozens of webpages. In our technique, we do not select the webpages randomly, we use a method to identify the webpages linked by the main menu of the website because they very likely contain the template. 
We only need to explore a few webpages to identify the webpages that implement the template.
Moreover, contrarily to us, they assume that all webpages in the website share the same template, and this is a strong limitation for many websites.

\section{Preliminaries} \label{sec_prelim}

The Document Object Model (DOM) \cite{DOM} is an API that provides programmers with a standard set of objects for the representation of HTML and XML documents. Our technique is based on the use of DOM as the model for representing webpages. Given a webpage, it is completely automatic to produce its associated DOM structure and vice-versa. In fact, current browsers automatically produce the DOM structure of all loaded webpages before they are processed.

The DOM structure of a given webpage is a tree where all the elements of the webpage are represented (included scripts and CSS styles) hierarchically. This means that a table that contains another table is represented with a node with a child that represents the internal table. 

\begin{figure}[h!]
	\centering
		\includegraphics[width=0.55\textwidth]{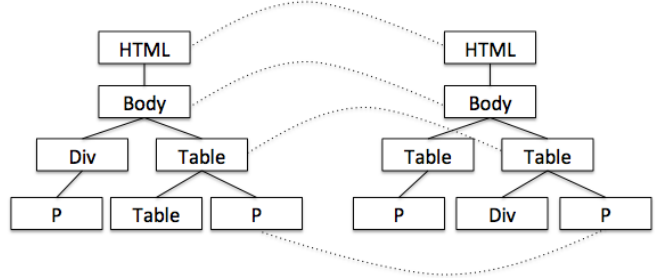}
	\caption{Equal top-down mapping between DOM trees}
	\label{fig-mapping}
\end{figure}

In the following, webpages are represented with a DOM tree $T = (N, E)$ where $N$ is a finite set of nodes and $E$ is a set of edges between nodes in $N$  
(see Figure~\ref{fig-mapping}). 
$root(T)$ denotes the root node of $T$. 
Given a node $n \in N$, 
$link(n)$ denotes the hyperlink of $n$ when $n$ is a node that represents a hyperlink (HTML label {\tt <a>}). 
$\mathit{parent}(n)$ represents node $n' \in N$ such that $(n', n)\in E$. 
Similarly, 
$\mathit{children}(n)$ represents the set $\{ n' \in N \mid (n, n') \in E \}$.
$subtree(n)$ denotes the subtree of $T$ whose root is $n \in N$. 
$path(n)$ is a non-empty sequence of nodes that represents a \emph{DOM path}; it can be defined as $path(n) = n_0 n_1 \dots n_m$ such that $\forall i, 0 \leq i < m . ~n_i = \mathit{parent}(n_{i + 1})$.

In order to identify the part of the DOM tree that is common in a set of webpages, our technique uses an algorithm that is based on the notion of mapping. A mapping establishes a correspondence between the nodes of two trees.


\begin{definition}{(based on Kuo's definition of mapping \cite{Tai79})}
\label{def_mapping}
A \emph{mapping} from a tree $T = (N, E)$ to a tree $T' = (N', E')$ is any set $M$ of pairs of nodes 
$(n, n') \in M$, $n \in N, n' \in N'$ such that, for any two pairs $(n_1, n'_1)$ and $(n_2, n'_2) \in M$, 
$n_1 = n_2$ iff $n'_1 = n'_2$.
\end{definition}

In order to identify templates, we are interested in a very specific kind of mapping that we call \emph{equal top-down mapping} (ETDM). 

\begin{definition}
\label{def_TDRmapping}
Given an equality relation $\triangleq$ between tree nodes,  
a mapping $M$ between two trees $T$ and $T'$ is said to be \emph{equal top-down} if and only if 
\begin{itemize}
\item {equal}: for every pair $(n, n') \in M$, $n \triangleq n'$. 
\item {top-down}: for every pair $(n, n') \in M$, with $n \neq root(T)$ and $n' \neq root(T')$, there is also a pair $(parent(n), parent(n')) \in M$.
\end{itemize}
\end{definition}

%



Note that this definition is parametric with respect to the equality function $\triangleq~\!\!\!\!\!: N \times N' \rightarrow [0..1]$ where $N$ and $N'$ are sets of nodes of two (often different) DOM trees. 
We could simply use the standard equality (=), but we left this relation open, to be general enough as to cover any possible implementation. In particular, other techniques consider that two nodes $n_1$ and $n_2$ are equal if they have the same label.
However, in our implementation we use a notion of node equality much more complex that uses the label of the node, its classname, its HTML attributes, its children, its position in the DOM tree, etc. 
  
This definition of mapping allows us to be more restrictive than other mappings such as, e.g., the \emph{restricted top-down mapping} (RTDM) introduced in \cite{ReiGSL04}. While RTDM permits the mapping of different nodes (e.g., a node labelled with \emph{table} with a node labelled with \emph{div}), ETDM can force all pairwise mapped nodes to have the same label. 
Figure~\ref{fig-mapping} shows an example of an ETDM using: 
$n \triangleq n'$ if and only if $n$ and $n'$ have the same label.
We can now give a definition of template using ETDM.

\begin{definition}
\label{def_template}
Let $p_0$ be a webpage whose associated DOM tree is $T_0=(N_0,E_0)$, and let $P=\{p_1\dots p_n\}$ be a collection of webpages with associated DOM trees $\{T_1\dots T_n\}$.
A \emph{template} of $p_0$ with respect to $P$ is a tree $(N,E)$ where 
\begin{itemize}
\item {nodes}: $N = \{n \in N_0 ~|~ \forall i, 1\leq i \leq n ~.~ (n,\_) \in M_{T_0,T_i}\}$ 
where $M_{T_0,T_i}$ is an equal top-down mapping between trees $T_0$ and $T_i$.
\item {edges}:   $ E= \{(m,m') \in E_0  ~|~ m,m' \in N \}$.
\end{itemize}
\end{definition}

Hence, the template of a webpage is computed with respect to a set of webpages (usually webpages in the same website). We formalize the template as a new webpage 
computed with an ETDM between the initial webpage and all the other webpages. 

\section{Template extraction} \label{sec_temex}

\begin{figure}[t]
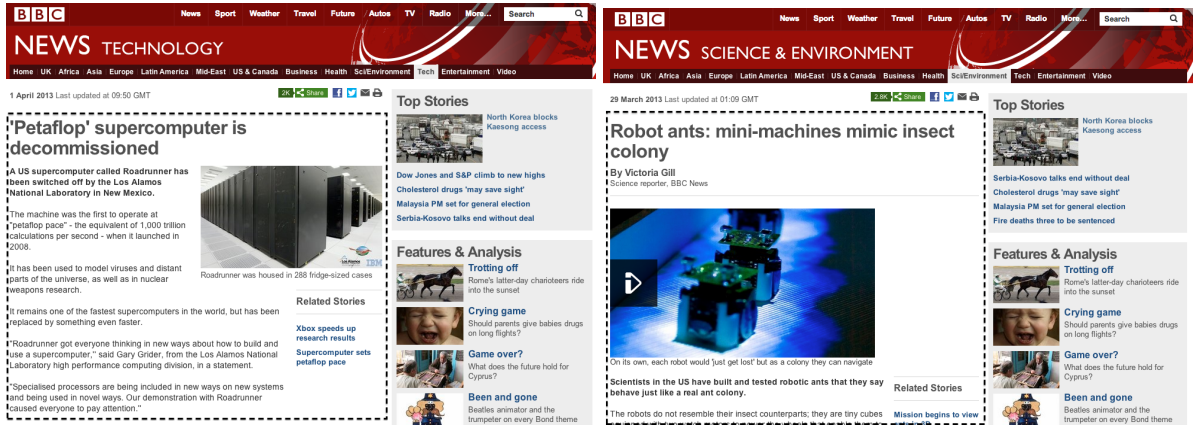

	\centering
		\includegraphics[width=0.49\textwidth]{\figures{pagelets1n.png}}
		\includegraphics[width=0.49\textwidth]{\figures{pagelets2n.png}}
	\caption{Webpages of BBC sharing a template}
	\label{fig-pagelets}
\end{figure}

Templates are often composed of a set of pagelets. Two of the most important pagelets in a webpage are the menu and the main content. For instance, 
in Figure~\ref{fig-pagelets} we see two webpages that belong to the ``News" portal of BBC. At the top of the webpages we find the main menu containing links to all BBC portals. We can also see a submenu under the big word ``News". The left webpage belongs to the ``Technology" section, while the right webpage belongs to the ``Science \& Environment" section. Both share the same menu, submenu, and general structure. 
In both pages the news are inside the pagelet in the dashed square. 
Note that this pagelet contains the main content and, thus, it should be indexed with a special treatment.
In addition to the main content, there is a common pagelet called ``Top Stories" with the most relevant news, and another one called ``Features and Analysis".

Our technique inputs a webpage (called key page) and it outputs its template. 
To infer the template, it analyzes some webpages from the (usually huge) universe of directly or indirectly linked webpages.
Therefore, we need to decide what concrete webpages should be analyzed. 
Our approach is very simple yet powerful: 
\begin{enumerate}
\item Starting from the key page, it identifies a complete subdigraph in the website topology, and then 
\item it extracts the template by calculating an ETDM between the DOM tree of the key page and some of the DOM trees of the webpages in the complete subdigraph. 
\end{enumerate}

Both processes are explained in the following sections.

\subsection{Finding a complete subdigraph in a website topology}

Given a website topology, a complete subdigraph (CS) represents a collection of webpages that are pairwise mutually linked. A n-complete subdigraph (n-CS) is formed by n nodes.  
Our interest in complete subdigraphs comes from the observation that the webpages linked by the items in a menu usually form a CS. This is a new way of identifying the webpages that contain the menu. At the same time, these webpages are the roots of the sections linked by the menu. 
The following example illustrates why menus provide very useful information about the interconnection of webpages in a given website. 

\begin{example}
Consider the BBC website. Two of its webpages are shown in Figure~\ref{fig-pagelets}. 
In this website all webpages share the same template, and this template has a main menu that is present in all webpages, and a submenu for each item in the main menu. 
The site map of the BBC website may be represented with the topology shown in Figure~\ref{fig-topology}. 

\begin{figure}[h]
	\centering
		\includegraphics[width=0.49\textwidth]{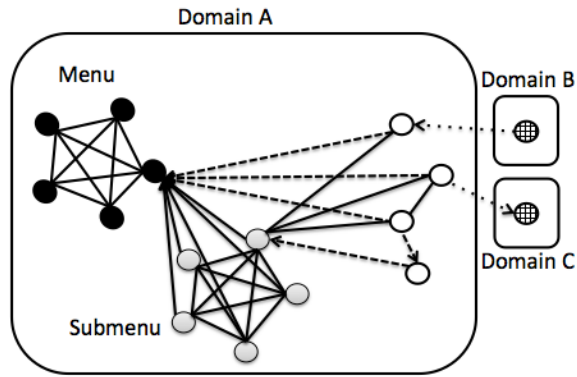}
	\caption{BBC Website topology}
	\label{fig-topology}
\end{figure}

In this figure, each node represents a webpage and each edge represents a link between two webpages. 
Solid edges are bidirectional, and dashed and dotted edges are directed. 
Black nodes are the webpages pointed by the main menu.
Because the main menu is present in all webpages, then all nodes are connected to all black nodes (we only draw some of the edges for clarity).   
Therefore all black nodes together form a complete graph (i.e., there is an edge between each pair of nodes). 
Grey nodes are the webpages pointed by a submenu, thus, all grey nodes together also form a complete graph. White nodes are webpages inside one of the categories of the submenu, therefore, all of them have a link to all black and all grey nodes.
\end{example}

Of course, not all webpages in a website implement the same template, and some of them only implement a subset of a template. For this reason, one of the main problems of template extraction is deciding what webpages should be analyzed. Minimizing the number of webpages analyzed is essential to reduce the web crawlers work. 
In our technique we introduce a new idea to select the webpages that must be analyzed: we identify a menu in the key page and we analyze the webpages pointed out by this menu. Observe that we only need to investigate the webpages linked by the key page, because they will for sure contain a CS that represents the menu.

In order to increase precision, we search for a CS that contains enough webpages that implement the template. This CS can be identified with Algorithm~\ref{AlgoMCS}. 

\begin{algorithm}[h!]
\caption{Extract a n-CS from a website}
\label{AlgoMCS}

\begin{algorithmic}
{\scriptsize
\smallskip
\STATE \textbf{Input:} An $initialLink$ that points to a webpage and the expected size $n$ of the CS.
\STATE \textbf{Output:} A set of links to webpages that together form a n-CS.\\ 
$~~~~~~~~~~~~~$If a n-CS cannot be formed, then they form the biggest m-CS with m $<$ n.
\medskip

\STATE $\textbf{begin}$
\STATE $~~\mathit{keyPage} = $ $\mathit{loadWebPage}$$(\mathit{initialLink})$;
\STATE $~~\mathit{reachableLinks} =  \mathit{getLinks} (\mathit{keyPage})$; 
\STATE $~~\mathit{processedLinks} = \emptyset$;
\STATE $~~\mathit{connections} = \emptyset$;
\STATE $~~\mathit{bestCS} = \emptyset$;
\STATE ~~\textbf{foreach} $link$ \textbf{in} $\mathit{reachableLinks}$
\STATE $~~~~~~\mathit{webPage} =  \mathit{loadWebPage} (link)$;
\STATE $~~~~~~\mathit{existingLinks} =  \mathit{getLinks} (webPage) \cap \mathit{reachableLinks}$;
\STATE $~~~~~~\mathit{processedLinks} = \mathit{processedLinks} \cup \{ link \}$;
\STATE $~~~~~~\mathit{connections} = \mathit{connections} \cup \{(link \rightarrow \mathit{existingLink}) \mid \mathit{existingLink} \in \mathit{existingLinks}\}$; 
\STATE $~~~~~~CS = \{ ls \in \cP(\mathit{processedLinks}) \mid link \in ls \land \forall l, l' \in ls~.~(l \rightarrow l'),(l' \rightarrow l) \in \mathit{connections} \}$;
\STATE $~~~~~~\mathit{maximalCS} = cs \in CS$ such that $\forall cs' \in CS~.~|cs| \geq |cs'|$;
\STATE ~~~~~~\textbf{if} $|\mathit{maximalCS}| = n$ \textbf{then} \textbf{return} $\mathit{maximalCS}$;
\STATE ~~~~~~\textbf{if} $|\mathit{maximalCS}| > |\mathit{bestCS|}$ \textbf{then} $\mathit{bestCS} = \mathit{maximalCS}$;
\STATE ~~\textbf{return} $\mathit{bestCS}$;
\STATE $\textbf{end}$


}
\end{algorithmic}
\end{algorithm}

This algorithm inputs a webpage and the size $n$ of the CS to be computed. We have empirically approximated the optimal value for $n$, which is 4.
The algorithm uses two trivial functions:
$\mathit{loadWebPage}(link)$, which loads and returns the webpage pointed by the input link, and 
$\mathit{getLinks}(webpage)$, which returns the collection of (non-repeated) links\footnote{In our implementation, this function removes those links that point to other domains because they are very unlikely to contain the same template.} in the input webpage (ignoring self-links). 
Observe that the main loop iteratively explores the links of the webpage pointed by the $initialLink$ (i.e., the key page) until it founds a n-CS. Note also that it only loads those webpages needed to find the n-CS, and it stops when the n-CS has been found. 
We want to highlight the mathematical expression
\begin{center}
\scriptsize
$CS = \{ ls \in \cP(processedLinks) \mid link \in ls \land \forall l, l' \in ls~.~(l \rightarrow l'),(l' \rightarrow l) \in connections \}$,
\end{center}
\vspace{-0.5em}
\noindent where $\cP(X)$ returns all possible partitions of set $X$.

It is used to find the set of all CS that can be constructed with the current $link$.
Here, $processedLinks$ contains the set of links that have been already explored by the algorithm. 
And $connections$ is the set of all links between the webpages pointed by $processedLinks$. 
$connections$ is used to identify the CS. 
Therefore, the set $CS$ is composed of the subset of $processedLinks$ that form a CS using $connections$.  

Observe that the current link must be part of the CS ($link \in ls$) to ensure that we make progress (not repeating the same search of the previous iteration). 
Moreover, because the CS is constructed incrementally, the statement
\begin{center}
\footnotesize
\textbf{if} $|\mathit{maximalCS}| = n$ \textbf{then} \textbf{return} $\mathit{maximalCS}$
\end{center}
\noindent ensures that whenever a n-CS can be formed, it is returned.    

\subsection{Template extraction from a complete subdigraph}

After we have found a set of webpages mutually linked and linked by the menu of the site (the complete subdigraph), 
we identify an ETDM between the key page and all webpages in the set. 
For this, initially, the template is considered to be the key page. Then, we compute an ETDM between the template and one webpage in the set. The result is the new refined template, that is further refined with another ETDM with another webpage, and so on until all webpages have been processed.
This process is formalized in Algorithm~\ref{Alg_template}, that uses function $\mathit{ETDM}$ to compute the biggest ETDM between two trees.

\begin{algorithm}[h!]
\caption{Extract a template from a set of webpages} \label{Alg_template}

\begin{algorithmic}
{\scriptsize
\smallskip
\STATE \textbf{Input:} A key page $p_k=(N_1,E_1)$ and a set of $n$ webpages $P$.
\STATE \textbf{Output:} A template for $p_k$ with respect to $P$.

\medskip

\STATE $\textbf{begin}$
\STATE $~~~template = p_k$;
\STATE ~~~\textbf{foreach} ($p$ \textbf{in} $P$)
\STATE ~~~~~~~~\textbf{if} $root(p_k) \triangleq root(p)$
\STATE $~~~~~~~~~~~~~template = \mathit{ETDM}(template,p)$; 
\STATE ~~~\textbf{return} $template$;
\STATE $\textbf{end}$

\medskip

\STATE \textbf{function} $\mathit{ETDM}$(tree $T_1 = (N_1,E_1)$, tree $T_2 = (N_2,E_2)$) 
\STATE ~~~$r_1 = root(T_1)$;
\STATE ~~~$r_2 = root(T_2)$;
\STATE ~~~nodes = $\{r_1\}$;
\STATE ~~~edges = $\emptyset$;
\STATE ~~~\textbf{foreach} $n_1 \in N_1\backslash nodes$, $n_2 \in N_2\backslash nodes~.~n_1 \underset{max}{\triangleq} n_2, (r_1,n_1)\in E_1$ and $(r_2,n_2)\in E_2$
\STATE ~~~~~~~~$(nodes\_st,edges\_st) = \mathit{ETDM}(subtree(n_1),subtree(n_2))$;
\STATE ~~~~~~~~$nodes = nodes \cup nodes\_st$;
\STATE ~~~~~~~~$edges = edges \cup edges\_st \cup \{(r_1,n_1)\} $;
\STATE ~~~\textbf{return} $(nodes,edges)$;
\STATE ~~~
}
\end{algorithmic}
\end{algorithm}

As in Definition~\ref{def_TDRmapping}, we left the algorithm parametric with respect to the equality function $\triangleq$.  
This is done on purpose, because this relation is the only parameter that is subjective and thus, it is a good design decision to leave it open. For instance, a researcher can decide that two DOM nodes are equal if they have the same label and attributes. Another researcher can relax this restriction ignoring some attributes (i.e, the template can be the same, even if there are differences in colors, sizes, or even positions of elements. It usually depends on the particular use of the extracted template). Clearly, $\triangleq$ has a direct influence on the precision and recall of the technique. The more restrictive, the more precision (and less recall). Note also that the algorithm uses $n_1 \underset{max}{\triangleq} n_2$ to indicate that $n_1$ and $n_2$ maximize function $\triangleq$. 

In our implementation, function $\triangleq$ is defined with a ponderation that compares two nodes considering their classname, their number of children, their relative position in the DOM tree, and their HTML attributes. We refer  the interested reader to our open and free implementation (\url{http://www.dsic.upv.es/~jsilva/retrieval/templates/}) where function $\triangleq$ is specified. 


\section{Implementation} \label{sec_impl}
The technique presented in this paper, including
all the algorithms, has been implemented as a Firefox's plugin. 
In this tool, the user can browse on the Internet as usual. 
Then, when he/she wants to extract the template of a webpage, he/she only needs to press the ``Extract Template'' button and the tool automatically 
loads the appropriate linked webpages to form a CS, analyzes them, and extracts the template. 
The template is then displayed in the browser as any other webpage. 
For instance, the template extracted for the webpages in Figure~\ref{fig-pagelets} contains the whole webpage except for the part inside the dashed box.

Our implementation and all the experimentation is public. 
All the information of the experiments, the source code of
the benchmarks, the source code of the tool,
and other material can be found at:

\begin{center}
\url{http://www.dsic.upv.es/~jsilva/retrieval/templates/}
\end{center}


\section{Conclusions}
\label{sec_concl}

Web templates are an important tool for website developers. 
By automatically inserting content into web templates, website developers, and content providers of large
web portals achieve high levels of productivity, and they produce webpages that are more usable thanks to their uniformity.

This work presents a new technique for template extraction. 
The technique is useful for website developers because they can automatically extract a clean HTML template of any webpage. This is particularly interesting to reuse components of other webpages.
Moreover, the technique can be used by other systems and tools such as indexers or wrappers as a preliminary stage.
Extracting the template allows them to identify the structure of the webpage and the topology of the website by analyzing the navigational information of the template. In addition, the template is useful to identify pagelets, repeated advertisement panels, and what is particularly important, the main content. 

Our technique uses the menus of a website to identify a set of webpages that share the same template with a high probability. 
Then, it uses the DOM structure of the webpages to identify the blocks that are common to all of them. 
These blocks together form the template.
To the best of our knowledge, the idea of using the menus to locate the template is new, and it allows us to quickly find a set of webpages from which we can extract the template. This is especially interesting for performance, because loading webpages to be analyzed is expensive, and this part of the process is minimized in our technique. As an average, our technique only loads 7 pages to extract the template. 

This technique could be also used for content extraction. Detecting the template of a webpage is very helpful to detect the main content. Firstly, the main content must be formed by DOM nodes that do not belong to the template. Secondly, the main content is usually inside one of the pagelets that are more centered and visible, and with a higher concentration of text. 
 
For future work, we plan to investigate a strategy to further reduce the amount of webpages loaded with our technique. 
The idea is to directly identify the menu in the key page by measuring the density of links in its DOM tree. 
The menu has probably one of the higher densities of links in a webpage. 
Therefore, our technique could benefit from measuring the links--DOM nodes ratio to directly find the menu in the key page, and thus, a complete subdigraph in the website topology.


\section{Acknowledgements}
\label{sec_acks}
This work has been partially supported by the EU (FEDER) and the
Spanish \emph{Ministerio de Econom\'{\i}a y Competitividad
(Secretar\'{\i}a de Estado de Investigaci\'on, Desarrollo e Innovaci\'on)}
under Grant TIN2013-44742-C4-1-R and by the
\emph{Generalitat Valenciana} under Grant PROMETEO/2011/052.
David Insa was partially supported
by the Spanish {Ministerio de Educaci\'on} under FPU Grant AP2010-4415.
Salvador Tamarit was partially supported by research project POLCA, Programming
Large Scale Heterogeneous Infrastructures (610686), funded by the
European Union, STREP FP7.

\bibliography{\biblio{biblio}}

\begin{thebibliography}{10}
\providecommand{\bibitemdeclare}[2]{}
\providecommand{\surnamestart}{}
\providecommand{\surnameend}{}
\providecommand{\urlprefix}{Available at }
\providecommand{\url}[1]{\texttt{#1}}
\providecommand{\href}[2]{\texttt{#2}}
\providecommand{\urlalt}[2]{\href{#1}{#2}}
\providecommand{\doi}[1]{doi:\urlalt{http://dx.doi.org/#1}{#1}}
\providecommand{\bibinfo}[2]{#2}

\bibitemdeclare{inproceedings}{BarR02}
\bibitem{BarR02}
\bibinfo{author}{Ziv \surnamestart Bar-Yossef\surnameend} \&
  \bibinfo{author}{Sridhar \surnamestart Rajagopalan\surnameend}
  (\bibinfo{year}{2002}): \emph{\bibinfo{title}{{T}emplate detection via data
  mining and its applications}}.
\newblock In: {\sl \bibinfo{booktitle}{{P}roceedings of the 11th
  {I}nternational {C}onference on {W}orld {W}ide {W}eb ({WWW}'02)}},
  \bibinfo{publisher}{{ACM}}, \bibinfo{address}{{N}ew {Y}ork, {NY}, {USA}}, pp.
  \bibinfo{pages}{580--591}, \doi{10.1145/511446.511522}.

\bibitemdeclare{inproceedings}{BarCKS08}
\bibitem{BarCKS08}
\bibinfo{author}{Marco \surnamestart Baroni\surnameend},
  \bibinfo{author}{Francis \surnamestart Chantree\surnameend},
  \bibinfo{author}{Adam \surnamestart Kilgarriff\surnameend} \&
  \bibinfo{author}{Serge \surnamestart Sharoff\surnameend}
  (\bibinfo{year}{2008}): \emph{\bibinfo{title}{{C}leaneval: a {C}ompetition
  for {C}leaning {W}eb {P}ages}}.
\newblock In: {\sl \bibinfo{booktitle}{{P}roceedings of the {I}nternational
  {C}onference on {L}anguage {R}esources and {E}valuation ({LREC}'08)}},
  \bibinfo{publisher}{{E}uropean {L}anguage {R}esources {A}ssociation}, pp.
  \bibinfo{pages}{638--643}.
\newblock
  \urlprefix\url{http://www.lrec-conf.org/proceedings/lrec2008/summaries/162.html}.

\bibitemdeclare{inproceedings}{BurR09}
\bibitem{BurR09}
\bibinfo{author}{Radek \surnamestart Burget\surnameend} \&
  \bibinfo{author}{Ivana \surnamestart Rudolfova\surnameend}
  (\bibinfo{year}{2009}): \emph{\bibinfo{title}{{W}eb {P}age {E}lement
  {C}lassification {B}ased on {V}isual {F}eatures}}.
\newblock In: {\sl \bibinfo{booktitle}{{P}roceedings of the 1st {A}sian
  {C}onference on {I}ntelligent {I}nformation and {D}atabase {S}ystems
  ({ACIIDS}'09)}}, \bibinfo{publisher}{{IEEE} {C}omputer {S}ociety},
  \bibinfo{address}{{W}ashington, {DC}, {USA}}, pp. \bibinfo{pages}{67--72},
  \doi{10.1109/ACIIDS.2009.71}.

\bibitemdeclare{inproceedings}{CarJLRC11}
\bibitem{CarJLRC11}
\bibinfo{author}{Eduardo \surnamestart Cardoso\surnameend},
  \bibinfo{author}{Iam \surnamestart Jabour\surnameend},
  \bibinfo{author}{Eduardo \surnamestart Laber\surnameend},
  \bibinfo{author}{Rog\'erio \surnamestart Rodrigues\surnameend} \&
  \bibinfo{author}{Pedro \surnamestart Cardoso\surnameend}
  (\bibinfo{year}{2011}): \emph{\bibinfo{title}{{A}n efficient
  language-independent method to extract content from news webpages}}.
\newblock In: {\sl \bibinfo{booktitle}{{P}roceedings of the 11th {ACM}
  symposium on {D}ocument {E}ngineering ({D}oc{E}ng'11)}},
  \bibinfo{publisher}{{ACM}}, \bibinfo{address}{{N}ew {Y}ork, {NY}, {USA}}, pp.
  \bibinfo{pages}{121--128}, \doi{10.1145/2034691.2034720}.

\bibitemdeclare{inproceedings}{Cha01}
\bibitem{Cha01}
\bibinfo{author}{Soumen \surnamestart Chakrabarti\surnameend}
  (\bibinfo{year}{2001}): \emph{\bibinfo{title}{{I}ntegrating the {D}ocument
  {O}bject {M}odel with hyperlinks for enhanced topic distillation and
  information extraction}}.
\newblock In: {\sl \bibinfo{booktitle}{{P}roceedings of the 10th
  {I}nternational {C}onference on {W}orld {W}ide {W}eb ({WWW}'01)}},
  \bibinfo{publisher}{{ACM}}, \bibinfo{address}{{N}ew {Y}ork, {NY}, {USA}}, pp.
  \bibinfo{pages}{211--220}, \doi{10.1145/371920.372054}.

\bibitemdeclare{misc}{DOM}
\bibitem{DOM}
\bibinfo{author}{{W3C} \surnamestart {C}onsortium\surnameend}
  (\bibinfo{year}{1997}): \emph{\bibinfo{title}{{D}ocument {O}bject {M}odel
  ({DOM})}}.
\newblock \bibinfo{howpublished}{{A}vailable from {URL}:
  \url{http://www.w3.org/{DOM}/}}.

\bibitemdeclare{inproceedings}{FerZBB08}
\bibitem{FerZBB08}
\bibinfo{author}{Adriano \surnamestart Ferraresi\surnameend},
  \bibinfo{author}{Eros \surnamestart Zanchetta\surnameend},
  \bibinfo{author}{Marco \surnamestart Baroni\surnameend} \&
  \bibinfo{author}{Silvia \surnamestart Bernardini\surnameend}
  (\bibinfo{year}{2008}): \emph{\bibinfo{title}{{I}ntroducing and evaluating
  {ukWaC}, a very large web-derived corpus of english}}.
\newblock In: {\sl \bibinfo{booktitle}{{P}roceedings of the 4th {W}eb as
  {C}orpus {W}orkshop ({WAC}-4)}}, pp. \bibinfo{pages}{47--54}.

\bibitemdeclare{inproceedings}{GibPT05}
\bibitem{GibPT05}
\bibinfo{author}{David \surnamestart Gibson\surnameend}, \bibinfo{author}{Kunal
  \surnamestart Punera\surnameend} \& \bibinfo{author}{Andrew \surnamestart
  Tomkins\surnameend} (\bibinfo{year}{2005}): \emph{\bibinfo{title}{{T}he
  volume and evolution of web page templates}}.
\newblock In \bibinfo{editor}{Allan \surnamestart Ellis\surnameend} \&
  \bibinfo{editor}{Tatsuya \surnamestart Hagino\surnameend}, editors: {\sl
  \bibinfo{booktitle}{{P}roceedings of the 14th {I}nternational {C}onference on
  {W}orld {W}ide {W}eb ({WWW}'05)}}, \bibinfo{publisher}{{ACM}}, pp.
  \bibinfo{pages}{830--839}, \doi{10.1145/1062745.1062763}.

\bibitemdeclare{inproceedings}{Got08}
\bibitem{Got08}
\bibinfo{author}{Thomas \surnamestart Gottron\surnameend}
  (\bibinfo{year}{2008}): \emph{\bibinfo{title}{{C}ontent {C}ode {B}lurring:
  {A} {N}ew {A}pproach to {C}ontent {E}xtraction}}.
\newblock In \bibinfo{editor}{A.~Min \surnamestart Tjoa\surnameend} \&
  \bibinfo{editor}{Roland~R. \surnamestart Wagner\surnameend}, editors: {\sl
  \bibinfo{booktitle}{{P}roceedings of the 19th {I}nternational {W}orkshop on
  {D}atabase and {E}xpert {S}ystems {A}pplications ({DEXA}'08)}},
  \bibinfo{publisher}{{IEEE} {C}omputer {S}ociety}, pp.
  \bibinfo{pages}{29--33}, \doi{10.1109/DEXA.2008.43}.

\bibitemdeclare{article}{InsST13}
\bibitem{InsST13}
\bibinfo{author}{David \surnamestart Insa\surnameend}, \bibinfo{author}{Josep
  \surnamestart Silva\surnameend} \& \bibinfo{author}{Salvador \surnamestart
  Tamarit\surnameend} (\bibinfo{year}{2013}): \emph{\bibinfo{title}{{U}sing the
  words/leafs ratio in the {DOM} tree for content extraction}}.
\newblock {\sl \bibinfo{journal}{{T}he {J}ournal of {L}ogic and {A}lgebraic
  {P}rogramming}} \bibinfo{volume}{82}(\bibinfo{number}{8}), pp.
  \bibinfo{pages}{311--325}, \doi{10.1016/j.jlap.2013.01.002}.

\bibitemdeclare{inproceedings}{Koh09}
\bibitem{Koh09}
\bibinfo{author}{Christian \surnamestart Kohlsch\"utter\surnameend}
  (\bibinfo{year}{2009}): \emph{\bibinfo{title}{{A} densitometric analysis of
  web template content}}.
\newblock In \bibinfo{editor}{Juan \surnamestart Quemada\surnameend},
  \bibinfo{editor}{Gonzalo \surnamestart Le\'on\surnameend},
  \bibinfo{editor}{Yo\"elle~S. \surnamestart Maarek\surnameend} \&
  \bibinfo{editor}{Wolfgang \surnamestart Nejdl\surnameend}, editors: {\sl
  \bibinfo{booktitle}{{P}roceedings of the 18th {I}nternational {C}onference on
  {W}orld {W}ide {W}eb ({WWW}'09)}}, \bibinfo{publisher}{{ACM}}, pp.
  \bibinfo{pages}{1165--1166}, \doi{10.1145/1526709.1526909}.

\bibitemdeclare{inproceedings}{KohFN10}
\bibitem{KohFN10}
\bibinfo{author}{Christian \surnamestart Kohlsch\"utter\surnameend},
  \bibinfo{author}{Peter \surnamestart Fankhauser\surnameend} \&
  \bibinfo{author}{Wolfgang \surnamestart Nejdl\surnameend}
  (\bibinfo{year}{2010}): \emph{\bibinfo{title}{{B}oilerplate detection using
  shallow text features}}.
\newblock In \bibinfo{editor}{Brian~D. \surnamestart Davison\surnameend},
  \bibinfo{editor}{Torsten \surnamestart Suel\surnameend},
  \bibinfo{editor}{Nick \surnamestart Craswell\surnameend} \&
  \bibinfo{editor}{Bing \surnamestart Liu\surnameend}, editors: {\sl
  \bibinfo{booktitle}{{P}roceedings of the 3th {I}nternational {C}onference on
  {W}eb {S}earch and {W}eb {D}ata {M}ining ({WSDM}'10)}},
  \bibinfo{publisher}{{ACM}}, pp. \bibinfo{pages}{441--450},
  \doi{10.1145/1718487.1718542}.

\bibitemdeclare{inproceedings}{KohN08}
\bibitem{KohN08}
\bibinfo{author}{Christian \surnamestart Kohlsch\"utter\surnameend} \&
  \bibinfo{author}{Wolfgang \surnamestart Nejdl\surnameend}
  (\bibinfo{year}{2008}): \emph{\bibinfo{title}{{A} densitometric approach to
  web page segmentation}}.
\newblock In \bibinfo{editor}{James~G. \surnamestart Shanahan\surnameend},
  \bibinfo{editor}{Sihem \surnamestart Amer-Yahia\surnameend},
  \bibinfo{editor}{Ioana \surnamestart Manolescu\surnameend},
  \bibinfo{editor}{Yi~\surnamestart Zhang\surnameend},
  \bibinfo{editor}{David~A. \surnamestart Evans\surnameend},
  \bibinfo{editor}{Aleksander \surnamestart Kolcz\surnameend},
  \bibinfo{editor}{Key-Sun \surnamestart Choi\surnameend} \&
  \bibinfo{editor}{Abdur \surnamestart Chowdhury\surnameend}, editors: {\sl
  \bibinfo{booktitle}{{P}roceedings of the 17th {ACM} {C}onference on
  {I}nformation and {K}nowledge {M}anagement ({CIKM}'08)}},
  \bibinfo{publisher}{{ACM}}, pp. \bibinfo{pages}{1173--1182},
  \doi{10.1145/1458082.1458237}.

\bibitemdeclare{inproceedings}{ReiGSL04}
\bibitem{ReiGSL04}
\bibinfo{author}{Davi de~Castro \surnamestart Reis\surnameend},
  \bibinfo{author}{Paulo~Braz \surnamestart Golgher\surnameend},
  \bibinfo{author}{Altigran~Soares \surnamestart Silva\surnameend} \&
  \bibinfo{author}{Alberto Henrique~Frade \surnamestart Laender\surnameend}
  (\bibinfo{year}{2004}): \emph{\bibinfo{title}{{A}utomatic web news extraction
  using tree edit distance}}.
\newblock In: {\sl \bibinfo{booktitle}{{P}roceedings of the 13th
  {I}nternational {C}onference on {W}orld {W}ide {W}eb ({WWW}'04)}},
  \bibinfo{publisher}{{ACM}}, \bibinfo{address}{{N}ew {Y}ork, {NY}, {USA}}, pp.
  \bibinfo{pages}{502--511}, \doi{10.1145/988672.988740}.

\bibitemdeclare{article}{Tai79}
\bibitem{Tai79}
\bibinfo{author}{Kuo~Chung \surnamestart Tai\surnameend}
  (\bibinfo{year}{1979}): \emph{\bibinfo{title}{{T}he {T}ree-to-{T}ree
  {C}orrection {P}roblem}}.
\newblock {\sl \bibinfo{journal}{{J}ournal of the {ACM}}}
  \bibinfo{volume}{26}(\bibinfo{number}{3}), pp. \bibinfo{pages}{422--433},
  \doi{10.1145/322139.322143}.

\bibitemdeclare{inproceedings}{VieSPMCF06}
\bibitem{VieSPMCF06}
\bibinfo{author}{Karane \surnamestart Vieira\surnameend},
  \bibinfo{author}{Altigran~S. \surnamestart da~Silva\surnameend},
  \bibinfo{author}{Nick \surnamestart Pinto\surnameend},
  \bibinfo{author}{Edleno~S. \surnamestart de~Moura\surnameend},
  \bibinfo{author}{Jo\~{a}o M.~B. \surnamestart Cavalcanti\surnameend} \&
  \bibinfo{author}{Juliana \surnamestart Freire\surnameend}
  (\bibinfo{year}{2006}): \emph{\bibinfo{title}{{A} fast and robust method for
  web page template detection and removal}}.
\newblock In: {\sl \bibinfo{booktitle}{{P}roceedings of the 15th {ACM}
  {I}nternational {C}onference on {I}nformation and {K}nowledge {M}anagement
  ({CIKM}'06)}}, \bibinfo{publisher}{{ACM}}, \bibinfo{address}{{N}ew {Y}ork,
  {NY}, {USA}}, pp. \bibinfo{pages}{258--267}, \doi{10.1145/1183614.1183654}.

\bibitemdeclare{inproceedings}{WenHH10}
\bibitem{WenHH10}
\bibinfo{author}{Tim \surnamestart Weninger\surnameend},
  \bibinfo{author}{William \surnamestart Henry~Hsu\surnameend} \&
  \bibinfo{author}{Jiawei \surnamestart Han\surnameend} (\bibinfo{year}{2010}):
  \emph{\bibinfo{title}{{CETR}: {C}ontent {E}xtraction via {T}ag {R}atios}}.
\newblock In \bibinfo{editor}{Michael \surnamestart Rappa\surnameend},
  \bibinfo{editor}{Paul \surnamestart Jones\surnameend},
  \bibinfo{editor}{Juliana \surnamestart Freire\surnameend} \&
  \bibinfo{editor}{Soumen \surnamestart Chakrabarti\surnameend}, editors: {\sl
  \bibinfo{booktitle}{{P}roceedings of the 19th {I}nternational {C}onference on
  {W}orld {W}ide {W}eb ({WWW}'10)}}, \bibinfo{publisher}{{ACM}}, pp.
  \bibinfo{pages}{971--980}, \doi{10.1145/1772690.1772789}.

\bibitemdeclare{inproceedings}{YiLL03}
\bibitem{YiLL03}
\bibinfo{author}{Lan \surnamestart Yi\surnameend}, \bibinfo{author}{Bing
  \surnamestart Liu\surnameend} \& \bibinfo{author}{Xiaoli \surnamestart
  Li\surnameend} (\bibinfo{year}{2003}): \emph{\bibinfo{title}{{E}liminating
  noisy information in Web pages for data mining}}.
\newblock In: {\sl \bibinfo{booktitle}{{P}roceedings of the 9th {ACM} {SIGKDD}
  {I}nternational {C}onference on {K}nowledge {D}iscovery and {D}ata mining
  ({KDD}'03)}}, \bibinfo{publisher}{{ACM}}, \bibinfo{address}{{N}ew {Y}ork,
  {NY}, {USA}}, pp. \bibinfo{pages}{296--305}, \doi{10.1145/956750.956785}.

\end{thebibliography}
\bibliographystyle{./eptcs}


\newpage



\end{document}